\magnification=\magstep1
\baselineskip=0.485cm
\hsize=16.4truecm \vsize=22.5truecm
\centerline{\bf  The $N$-soliton solution of the Degasperis-Procesi equation }\par
\bigskip
\centerline{Yoshimasa Matsuno
\footnote{$^{a)}$}{{\it E-mail address}: matsuno@yamaguchi-u.ac.jp}} \par 
\centerline{\it Department of Applied Science, Faculty of Engineering,}
\par
\centerline{\it Yamaguchi University, Ube 755-8611, Japan} \par
\bigskip
\noindent {\bf Abstract} \par
\noindent This paper extends the results of  the previous paper 
designated I hereafter in which the one-
and two-soliton solutions of the Degasperis-Procesi (DP)
equation were obtained and their peakon limit was considered. 
Here, we present the general $N$-soliton
solution of the DP equation and investigate its property. 
We show that  it has a novel
structure expressed by a parametric representation in terms of
 the BKP $\tau$-functions.
A purely algebraic proof of the solution is given by establishing various identities 
among the  $\tau$-functions.  The large time asymptotic of the solution  recovers the
formula for the phase shift which was derived in I by a different method. 
Finally, the structure of the $N$-soliton solution is discussed
in comparison with that of the Camassa-Holm shallow water wave equation. 
  \par
\vfill \eject
\leftline{\bf  1. Introduction}\par
\noindent The Degasperis-Procesi (DP) equation
$$u_t+3\kappa^3 u_x-u_{xxt}+4uu_x=3 u_xu_{xx}+uu_{xxx} \eqno(1.1)$$
where $u=u(x,t)$ and $\kappa$ is a real constant, is a current research interest in 
soliton theory. 
Here, the subscripts $t$ and $x$ appended to $u$ denote partial differentiation.
This equation has been proposed as  a candidate for the completely
integrable nonlinear partial differential equations 
by using the  method of asymptotic integrability [1].  A subsequent work has established
its complete integrability by exhibiting  some features common to the integrable system 
such as  the existence  of the Lax pair, infinity of  conservation laws
and so on [2].  Most studies so far have been devoted to 
the special case $\kappa=0$ in equation (1.1) [2-6].  
Among them, a remarkable 
feature is that  the corresponding equation  exhibits
peakon solutions which are represented by piecewise analytic functions. On the other
hand, the case $\kappa\not=0$  merits a separate treatment  and it
was considered quite recently in our work, which is designated I hereafter [7].
Indeed, performing  a reduction procedure for the soliton solutions of the 
Kadomtsev-Petviashvili (KP) hierarchy,
we have obtained the explicit one- and two-soliton solutions and explored 
their properties.  
We have found that as opposed to the case $\kappa=0$, solutions
recover their analytic nature and behave like usual solitons. 
Nevertheless, several new features appear which are worth remarking. First,
 the soliton velocity depends nonlinearly on its amplitude as
opposed to the usual linear relation.
Second, the interaction
process of two solitons reveals a new aspect which was never observed in the interaction of
solitons for the typical soliton equation like the Korteweg-de Vries (KdV) equation.
In particular, the small soliton    exhibits a nonnegative phase
shift for certain range of the  amplitude parameters which is in striking contrast to the KdV case
where  the small soliton always exhibits a negative phase shift irrespective of values
of the amplitude  parameters. \par
  The form 
of the two-soliton solution presented in I is very suggestive in 
constructing the general $N$-soliton solution where $N$ is an arbitrary 
positive integer. 
We emphasize that 
it has a parametric representation in terms of
two  $\tau$-functions each of which has the
standard Hirota form. This fact would enable us to surmise the functional form of the 
 $\tau$-functions which produce the $N$-soliton solution.  
Nevertheless, the gap between the two-soliton and $N$-soliton cases is found to be 
so great that the construction of the $N$-soliton solution becomes a nontrivial 
problem which deserves further study.
\par
The purpose of this paper is to provide the general $N$-soliton solution of the DP
equation (1.1)
in the form of
the parametric representation. 
The solution presented here will be shown to exhibit several new features when compared 
with existing $N$-soliton solutions of typical nonlinear evolution equations like the 
KdV and KP equations. 
The procedure for constructing the $N$-soliton solution can be accomplished along
the same line as in the previous work I. However, we encounter several technical
difficulties in the general $N$-soliton case which arise in reducing the proof 
of the solution to certain algebraic identities. Thus, the establishment of these identities
 turns out to be the main task in the present analysis. 
In conclusion, 
we show that as in the case of 
the two-soliton solution, the general $N$-soliton solution has a simple structure
expressed by two fundamental  $\tau$-functions.  Remarkably,  their functional form is 
found to be 
the same as the  $\tau$-function for the $N$-soliton solution of a model shallow 
water wave equation introduced by Hirota and Satsuma [8].  
In section 2, we summarize the results
obtained in  the  previous study  for constructing soliton solutions of the DP equation. 
In section 3, we carry out the proof of the $N$-soliton solution by means of 
a purely algebraic procedure
in which the 
identities among the  $\tau$-functions will play a central role. In section 4, 
we investigate the asymptotic 
behavior of the $N$-soliton solution for large time and recover the formula for the 
phase shift which has been derived in I by a different method. 
Section 5 is devoted to discussion where the structure of the $N$-soliton
solution is compared with that of the Camassa-Holm (CH) shallow water wave
equation.  In appendix, we apply the CKP reduction to the  $\tau$-function 
of the $2N$-soliton solution for the 
KP hierarchy and present the $\tau$-function which produces the $N$-soliton
solution for a member of the CKP hierarchy.  \par
\bigskip
\leftline{\bf 2. Summary of the previous study}\par
\leftline{\it 2.1. A system of equations equivalent to the DP equation} \par
\noindent Here, we summarize the results associated with the DP equation
 while focusing  on an equivalent system
of equations to the DP equation  and their soliton solutions. 
We describe only the main formulas without further explanation. For their
derivation and implication, we refer to our work I as well as a related paper [2]. \par
In order to solve equation (1.1)
under the boundary condition $u(\pm\infty, t)=0$, we introduce the new variable $r$ 
by the relation
$$r^3=u-u_{xx}+\kappa^3  \eqno(2.1)$$
and recast equation (1.1) into the conservation form
$$r_t+(ru)_x=0 \eqno(2.2)$$
where the boundary condition for $r$ is given by $r(\pm\infty, t)=\kappa$. 
A crucial step  for simplifying the analysis is to define a coordinate transformation
$(x, t)\rightarrow (y, \tau)$ by
$$dy=rdx-rudt  \eqno(2.3a)$$
$$ d\tau=dt.  \eqno(2.3b)$$
Consequently,  the $x$ and $t$ derivatives are rewritten as
$${\partial\over\partial x}=r{\partial\over\partial y}, \eqno(2.4a)$$
$${\partial\over\partial t}
={\partial\over\partial \tau}-ru{\partial\over\partial y}.\eqno(2.4b)$$
The  transformation (2.3) is well-defined provided that $r>0$. Under this condition, its 
inverse transformation is  obtained by solving the system of equations for $x$
$$x_y={1\over r(y,t)} \eqno(2.5a)$$
$$x_t=u(y, t). \eqno(2.5b)$$
Here in equation (2.5) and 
hereafter, we use the time variable $t$ instead of $\tau$ in view of (2.3b).
Using the transformation (2.3), equations (2.1) and (2.2) become
$$u=-r(\ln\ r)_{t y}+r^3-\kappa^3  \eqno(2.6)$$
$$r_t+r^2u_y=0.  \eqno(2.7)$$
Note that by virtue of (2.7) the compatibility condition $x_{yt}=x_{ty}$ is assured 
for the system of equations (2.5) that determine the inverse mapping. \par
\leftline{\it 2.2 Lax hierarchy of the CKP equation}\par
\noindent  To demonstrate the complete integrability of the  system of equations (2.6) and (2.7), we
substitute equation (2.6) into equation (2.7) to obtain
$$v_t=-{3\over 4}(r^2)_y  \eqno(2.8a)$$  
where the new dependent variable $v$ is defined by
$$v=-{r_{yy}\over 2r}+{r_y^2\over 4r^2}-{1\over 4}\left({1\over r^2}
-{1\over \kappa^2}\right). \eqno(2.8b)$$ 
Furthermore, we can eliminate the variable $r$ from (2.8) and derive a single
nonlinear equation for $v$. It reads 
$$v_{tyyyy}+\left(20v-{5\over\kappa^2}\right)v_{tyy}+30v_yv_{ty}+\left\{18v_{yy}
+16\left(2v-{1\over 2\kappa^2}\right)^2\right\}v_t$$
$$+\left\{4v_{yyy}
+32\left(2v-{1\over 2 \kappa^2}\right)v_y\right\}\left(\partial_y^{-1}v_t-{3\over 4}\kappa^2\right)=0  \eqno(2.9)$$
where $\partial^{-1}_y=\int^y_{-\infty}dy$ is an integral operator.
It has been pointed out that equation (2.9) with $v+(1/4\kappa^2)$ 
in place of $v$ is a member
of the Lax hierarchy of the Kaup-Kuperschmidt (KK) (or CKP) equation. Thanks to 
this fact, we were able to construct explicitly the $N$-soliton solution of equation (2.9). \par
\leftline{\it  2.3.  Procedure for constructing N-soliton solution} \par
\noindent   The procedure for constructing 
the $N$-soliton solution of the DP equation consists of
two steps, which we shall now demonstrate. 
 First, we write the $N$-soliton solution of equation (2.9) in the form
$$v={3\over 4}({\rm ln}\ f)_{yy} \eqno(2.10)$$
where $f$ is obtained by means of a reduction for the $2N$-soliton solution of the KP
hierarchy. See appendix  for detail. It may be expressed in the  form of finite sum
$$f=\sum_{\mu, \nu=0, 1}{\rm exp}\Bigg[\sum_{i=1}^N(\mu_i+\nu_i)
\xi_i+ \sum_{i=1}^N(2\mu_i\nu_i-\mu_i-\nu_i){\rm ln}\ a_i$$
$$+{1\over 2}\sum_{i,j=1\atop (i\not=j)}^N(\mu_i\mu_j+\nu_i\nu_j)A_{2i-1, 2j-1}
+{1\over 2}\sum_{i,j=1\atop (i\not=j)}^N(\mu_i\nu_j+\mu_j\nu_i)A_{2i-1, 2j}
\Bigg]  \eqno(2.11a)$$
with
$$\xi_i=k_i\left(y-{3\kappa^4\over 1-\kappa^2k_i^2}t-y_{i0}\right), (i=1, 2, ..., N)
\eqno(2.11b)$$
$$a_i=\sqrt{{1-{1\over 4}\kappa^2k_i^2\over 1-\kappa^2k_i^2}}, (i=1, 2, ..., N)
 \eqno(2.11c)$$
$${\rm e}^{A_{2i-1, 2j-1}}={(p_i-p_j)(q_i-q_j)\over  (p_i+q_j)(q_i+p_j)},
(i, j=1, 2, ..., N; i\not=j)  \eqno(2.11d)$$
  $${\rm e}^{A_{2i-1, 2j}}={(p_i-q_j)(q_i-p_j)\over  (p_i+p_j)(q_i+q_j)},
(i, j=1, 2, ..., N)  \eqno(2.11e)$$
where the parameters  $p_i$ and $q_i$ in (2.11d) and (2.11e) are given respectively by
$$p_i={k_i\over 2}
\left[1+{2\over \kappa k_i}\sqrt{{1\over 3}\left(1-{1\over 4}\kappa^2k_i^2\right)}\right],
(i=1, 2, ..., N) \eqno(2.11f)$$
$$q_i={k_i\over 2}
\left[1-{2\over \kappa k_i}\sqrt{{1\over 3}\left(1-{1\over 4}\kappa^2k_i^2\right)}\right],
 (i=1, 2, ..., N). \eqno(2.11g)$$
 Here, $k_i$ and $y_{i0}$ are the wavenumber and phase of the $i$th soliton, respectively and
the conditions $0<\kappa k_i<1, (i=1, 2, ..., N)$ are imposed 
to ensure the positivity of $r$. Thus, the $N$-soliton solution of equation (2.9)
is characterized completely by the $2N$ parameters $k_i$ and $y_{i0} (i=1, 2, ..., N)$.  
For $N=1, 2$,  the explicit expressions of $f$  are written as
$$f= 1+{2\over a_1}{\rm e}^{\xi_1}+{\rm e}^{2\xi_1},  (N=1)   \eqno(2.12a)$$
$$f=  1+{2\over a_1}{\rm e}^{\xi_1}
+{2\over a_2}{\rm e}^{\xi_2}+  {\rm e}^{2\xi_1}+{\rm e}^{2\xi_2}$$
$$+{2\nu\over a_1a_2}{\rm e}^{\xi_1+\xi_2}+{2\delta\over a_2}{\rm e}^{2\xi_1+\xi_2}
+{2\delta\over a_1}{\rm e}^{\xi_1+2\xi_2}
+\delta^2{\rm e}^{2\xi_1+2\xi_2},   (N=2) \eqno(2.12b)$$
where
$$\delta={(k_1-k_2)^2[(k_1^2-k_1k_2+k_2^2)\kappa^2-3]\over 
          (k_1+k_2)^2[(k_1^2+k_1k_2+k_2^2)\kappa^2-3]} \eqno(2.12c)$$
$$\nu={(2k_1^4-k_1^2k_2^2+2k_2^4)\kappa^2-6(k_1^2+k_2^2)\over
 (k_1+k_2)^2[(k_1^2+k_1k_2+k_2^2)\kappa^2-3]}.\eqno(2.12d)$$
\par
Second, we substitute (2.10) into (2.8a) and integrate the resultant equation by $y$ 
under the boundary condition $r(\pm\infty, t)=\kappa$ and obtain the expression of $r$
in terms of $f$                                 
$$r^2=-(\ln \ f)_{ty}+\kappa^2.\eqno(2.13)$$
To represent the solution in parametric form, we must integrate the system of equations (2.5).
It follows from equation (2.5a) 
subject  to  the boundary condition $r(\pm\infty, t)=\kappa$ that
$$x={y\over \kappa}+\int^y_{-\infty}\left({1\over r}-{1\over \kappa}\right)dy
+d  \eqno(2.14)$$
where $d$ is an integration constant. This constant is independent of $t$ as confirmed easily
using equations (2.5b) and  (2.7). Substituting  $r$ from (2.13) into (2.14) and  performing 
the integration, we obtain the expression of the coordinate transformation 
$x=x(y, t)$, which, combined with (2.5b), gives a parametric representation of the
$N$-soliton solution for the DP equation.  
In the simplest case of $N=1$,  the  explicit form of 
the one-soliton solution  reads
$$u(y,t)={{8\kappa^3\over a_1}(a_1^2-1)\left(a_1^2-{1\over 4}\right)
\over \cosh\ \xi_1 +2a_1-{1\over a_1}}  \eqno(2.15a)$$
$$x(y,t)={y\over \kappa}+{\rm ln}\left({\alpha_1+1+(\alpha_1-1){\rm e}^{\xi_1}\over
 \alpha_1-1+(\alpha_1+1){\rm e}^{\xi_1}}\right). \eqno(2.15b)$$
Here, the integration constant $d$ has been set to 
$d={\rm ln}[(\alpha_1+1)/(\alpha_1-1)]$  with $\alpha_1$ being expressed in terms of
$a_1$ as
$$\alpha_1=\sqrt{(2a_1-1)(a_1+1)\over (2a_1+1)(a_1-1)} \eqno(2.15c)$$
and $\xi_1$ and $a_1$ are defined respectively by (2.11b) and (2.11c).
\par
\leftline{\it 2.4. Remark}\par
\noindent The most difficult technical problem throughout  the present analysis is 
to evaluate the integral in (2.14)
with  $r$ constructed from the $N$-soliton  $\tau$-function $f$  through the 
relation (2.13). In I, this calculation has been performed only for the case of $N =1, 2.$
As will be developed in detail in the next section, the problem mentioned here is
resolved completely by finding a suitable expression of $r$ in terms of two fundamental
 $\tau$-functions. 
Another remark is concerned with the form of the  $\tau$-function $f$. 
As shown in appendix, it can be represented by an equivalent determinantal form which 
has been used in I. However, $f$ given by (2.11) is relevant in establishing  the
key identities (3.13) and (3.14) below since the  $\tau$-functions $g_1$ and $g_2$ 
cannot be put into the form of  determinants  but they
are expressed by pfaffians. Consequently, we can no longer rely on
the various formulas for determinants used frequently in proving the analogous
identities associated with the $N$-soliton  solution of the CH equation [9]. 
\par 
\bigskip
\leftline{\bf 3.  $N$-soliton solution}\par
\leftline{\it 3.1. Parametric representation}\par
\noindent Now, the main result  in this paper can be formulated as follows: 
The $N$-soliton solution of the  
DP equation (1.1) can be written compactly in a parametric representation
$$u(y,t)=\left({\rm ln}\ {g_1\over g_2}\right)_t  \eqno(3.1a)$$
$$x(y,t)={y\over\kappa} +{\rm ln}\ {g_1\over g_2} +d. \eqno(3.1b)$$
Here, $g_1$ and $g_2$ are given respectively by
$$g_1=\sum_{\mu=0,1}{\rm exp}\left[\sum_{i=1}^N\mu_i(\xi_i-\phi_i)
+\sum_{i,j=1\atop (i<j)}^N\mu_i\mu_j\gamma_{ij}\right]
\eqno(3.2a)$$
$$g_2=\sum_{\mu=0,1}{\rm exp}\left[\sum_{i=1}^N\mu_i(\xi_i+\phi_i)
+\sum_{i,j=1\atop (i<j)}^N\mu_i\mu_j\gamma_{ij}\right]
\eqno(3.2b)$$
with
$${\rm e}^{-\phi_i}=\sqrt{\left(1-{\kappa k_i\over 2}\right)\left(1-\kappa k_i\right)
\over \left(1+{\kappa k_i\over 2}\right)\left(1+\kappa k_i\right)}
(i=1, 2, ..., N) \eqno(3.2c)$$
$$\eqalignno{{\rm e}^{\gamma_{ij}} &=
{(p_i-p_j)(p_i-q_j)(q_i-p_j)(q_i-q_j)\over
 (p_i+p_j)(p_i+q_j)(q_i+p_j)(q_i+q_j)} \cr
&={(k_i-k_j)^2[(k_i^2-k_ik_j+k_j^2)\kappa^2-3]\over
(k_i+k_j)^2[(k_i^2+k_ik_j+k_j^2)\kappa^2-3]}, 
 (i, j=1, 2, ..., N; i\not=j)  &(3.2d)}$$
where the phase variable $\xi_i$ of the $i$th soliton is defined by (2.11b). 
In deriving the second line of (3.2d), we have used (2.11f) and (2.11g). 
Note remarkably that $g_1$ and $g_2$ have the same functional form except  for
the phase factors $\pm\phi_i$. \par
\leftline{\it 3.2.  Identities among $\tau$-fuctions}\par
\noindent We verify (3.1)  in a sequence of steps. To this end, we first define the following quantities 
constructed from the  $\tau$-function introduced in (A.1):
$$\tilde f=\tau_{2N}(\tilde\xi_1, \tilde\xi_2, ..., \tilde\xi_{2N}) \eqno(3.3)$$
$$\tilde f_1=\tau_{2N}(\tilde\xi_1+{\rm  ln}(\tilde p_1/\tilde q_1), 
\tilde\xi_2+{\rm  ln}(\tilde p_2/\tilde q_2), ..., \tilde\xi_{2N}+{\rm  ln}(\tilde p_{2N}/\tilde q_{2N})) \eqno(3.4)$$
$$\tilde f_2=\tau_{2N}(\tilde\xi_1+{\rm  ln}(\tilde q_1/\tilde p_1), 
\tilde\xi_2+{\rm  ln}(\tilde q_2/\tilde p_2), ..., \tilde\xi_{2N}+{\rm  ln}(\tilde q_{2N}/\tilde p_{2N})). \eqno(3.5)$$
An important observation here is that the following bilinear identity holds  among
$\tilde f, \tilde f_1$ and $\tilde f_2$:
$$\kappa^2\tilde f^2-\tilde f\tilde f_{ty}+\tilde f_t\tilde f_y=\kappa^2\tilde f_1\tilde f_2.
\eqno(3.6)$$
We have already met with
this type of identity  in a previous paper dealing with
the $N$-soliton solution of the CH equation [9].  
To modify (3.6) into a known identity, it is sufficient to replace the 
expressions of  $\tilde f, \tilde f_1$ and $\tilde f_2$
 given above by their determinantal forms using the relation [10]
$$\sum_{\mu=0,1}{\rm exp}\left[\sum_{i=1}^{2N}\mu_i\tilde\xi_i
+\sum_{i,j=1\atop (i<j)}^{2N}\mu_i\mu_j\tilde A_{ij}\right]
=\lambda_{2N}\ {\rm det}\left({{\rm e}^{\tilde\zeta_i}\delta_{ij}\over \tilde p_i-\tilde q_i}
+{1\over \tilde p_i-\tilde q_j}\right )_{1\leq i,j\leq 2N} \eqno(3.7a)$$
where
$$\tilde \zeta_i=\tilde\xi_i+\sum_{j=1\atop (j\not=i)}^{2N}\tilde A_{ij},
(i=1, 2, ..., 2N) \eqno(3.7b)$$
$\lambda_{2N}$ is a constant factor given by
$$\lambda_{2N}={\rm exp}\left[-\sum_{i,j=1\atop(i<j)}^{2N}\tilde A_{ij}\right]
\prod_{i=1}^{2N}(\tilde p_i-\tilde q_i) \eqno(3.7c)$$
and $\delta_{ij}$ is Kronecker's delta. Then, we can confirm that (3.6) reduces to
the bilinear identity (23) of  [9]. \par
Let us now apply the CKP reduction 
defined by (A.4) to $\tilde f, \tilde f_1$ and $\tilde f_2$
and  subsequently shift the phase variables 
as $\xi_i\rightarrow \xi_i-{\rm ln}\ a_i (i=1, 2, ..., N)$ to
introduce the new quantities $f$ and $g$ 
$$f=\tilde f|_{CKP}(\xi_i\rightarrow \xi_i-{\rm ln}\ a_i)  \eqno(3.8)$$
$$g=\tilde f_1|_{CKP}(\xi_i\rightarrow \xi_i-{\rm ln}\ a_i)
= \tilde f_2|_{CKP}(\xi_i\rightarrow \xi_i-{\rm ln}\ a_i).
\eqno(3.9)$$
Here, $g$ is represented by the finite sum analogous to the expression of $f$ 
$$g=\sum_{\mu, \nu=0, 1}{\rm exp}\Bigg[\sum_{i=1}^N(\mu_i+\nu_i)
\xi_i+ \sum_{i=1}^N(2\mu_i\nu_i-\mu_i-\nu_i){\rm ln}\ a_i
+\sum_{i=1}^N(\mu_i-\nu_i){\rm ln}\left(-{q_i\over p_i}\right)$$
$$+{1\over 2}\sum_{i,j=1\atop (i\not=j)}^N(\mu_i\mu_j+\nu_i\nu_j)A_{2i-1, 2j-1}
+{1\over 2}\sum_{i,j=1\atop (i\not=j)}^N(\mu_i\nu_j+\mu_j\nu_i)A_{2i-1, 2j}
\Bigg]. \eqno(3.10)$$
Note that only the difference between $f$ and $g$ is the third term
on the right-hand side of (3.10).
It is easy to see that both $\tilde f_1$ and $\tilde f_2$ reduce to the  expression of $g$ 
given above as a consequence of 
the CKP reduction. \par
It turns out from (3.6), (3.8) and (3.9) that
$$\kappa^2-({\rm ln}\ f)_{ty}=\kappa^2\left({g\over f}\right)^2. \eqno(3.11)$$
Comparing (3.11) with (2.13) and taking account of the boundary condition
$r(\pm\infty, t)=\kappa$, we obtain the simplified expression of $r$ in terms of $f$ and $g$
$$r=\kappa{g\over f}. \eqno(3.12)$$
Thus, $r$ becomes a rational function of ${\rm e}^{\xi_i} (i=1, 2, ..., N)$. This significant 
expression would not be  derived without noticing the identity (3.6) since $r$ from (2.13)
leads simply to an irrational function of ${\rm e}^{\xi_i} (i=1, 2, ..., N)$.  The compact
expression (3.12) enables us to integrate (2.14) explicitly when coupled with
the following set of  identities among $f, g, g_1$ and $g_2$:
$$f=g_1g_2+\kappa(g_{1,y}g_2-g_1g_{2,y})  \eqno(3.13)$$
$$g=g_1g_2. \eqno(3.14)$$
 The proof of the above identities will be done by mathematical induction similar to 
that used successfully for the proof of the $N$-soliton solutions 
of various soliton equations within the framework of the bilinear
formalism [11]. \par
\leftline{\it 3.2.1. Proof of (3.13)}\par
\noindent The identity (3.13) is established by  comparing the coefficient of the factor
${\rm exp}\Bigl[\sum_{i=1}^n\xi_i$ \par
\noindent $+\sum_{i=n+1}^m2\xi_i\Bigr] (0\leq n<m\leq N)$ on
both sides.  Let $L_{m, n}$ be the corresponding coefficient of $f$  and $R_{m, n}$ be 
the one for
the right-hand side of (3.13).  Correspondingly, the summation
 with respect to
$\mu_i$ and $\nu_i$ must be performed under the conditions
$$\mu_i+\nu_i=1, (i=1, 2, ..., n), \ \mu_i=\nu_i=1, (i=n+1, n+2, ..., m),$$
$$\mu_i=\nu_i=0, (i=m+1, m+2, ..., N). \eqno(3.15)$$
For the purpose of performing the proof effectively,
it is convenient to  introduce the new variables $\sigma_i$  according to the relations
$$\mu_i={1\over 2}(1+\sigma_i), \ \nu_i={1\over 2}(1-\sigma_i),
(i=1, 2, ..., N)  \eqno(3.16)$$
where $\sigma_i$ takes either the value $+1$ or $-1$.
Consequently
$$\mu_i\mu_j+\nu_i\nu_j={1\over 2}(1+\sigma_i\sigma_j), (i, j=1, 2, ..., N) 
 \eqno(3.17a)$$
$$\mu_i\nu_j+\mu_i\nu_i={1\over 2}(1-\sigma_i\sigma_j), (i, j=1, 2, ..., N). \eqno(3.17b)$$
Taking account of the relations which follow from (2.11d)-(2.11g) and (3.2d) 
$$A_{2i-1,2i}=2\ {\rm ln}\ a_i, (i=1,  2, ..., N)  \eqno(3.18a)$$
$$A_{2i-1,2j}+A_{2i,2j-1}=\gamma_{ij}, (i,j=1, 2, ..., N; i\not=j)  \eqno(3.18b)$$
as well as (3.15) and (3.17), we find that
$$L_{m,n}=\sum_{\sigma=\pm 1}{\rm exp}\Bigg[-\sum_{i=1}^n{\rm ln}\  a_i
+{1\over 4} \sum_{i,j=1\atop(i\not=j)}^n(1+\sigma_i\sigma_j)\gamma_{ij}$$
$$-{1\over 2} \sum_{i,j=1\atop(i\not=j)}^n\sigma_i\sigma_jA_{2i-1,2j}+\sum_{i=n+1}^m
\sum_{j=1\atop(j\not=i)}^m\gamma_{ij}\Bigg]  \eqno(3.19)$$
$$R_{m,n}=\sum_{\sigma=\pm 1}\left(1+\kappa\sum_{i=1}^n\sigma_ik_i\right)
{\rm exp}\Bigg[-\sum_{i=1}^n\sigma_i\phi_i
+{1\over 4} \sum_{i,j=1\atop(i\not=j)}^n(1+\sigma_i\sigma_j)\gamma_{ij}
+\sum_{i=n+1}^m
\sum_{j=1\atop(j\not=i)}^m\gamma_{ij}\Bigg]. \eqno(3.20)$$
To simplify (3.19) further, we use the relation
$${\rm exp}\left[{1\over 2}\{\gamma_{ij}
+\sigma_i\sigma_j(A_{2i-1,2j-1}-A_{2i-1,2j})\}\right]$$
$$\eqalignno{ & ={1\over 2}(1+\sigma_i\sigma_j){(p_i-p_j)(q_i-q_j)\over (p_i+q_j)(q_i+p_j)}
+{1\over 2}(1-\sigma_i\sigma_j){(p_i-q_j)(q_i-p_j)\over (p_i+p_j)(q_i+q_j)} \cr
&={1\over 2}{b_{ij}\over d_{ij}}+{1\over 2}\sigma_i\sigma_j{\beta_i\beta_j\over d_{ij}},
(i,j=1, 2, ..., N; i\not=j)
& (3.21a)}$$
where 
$$\beta_i=2\sqrt{3}k_i\sqrt{1-{1\over 4}\kappa^2k_i^2},
(i=1, 2, ..., N)  \eqno(3.21b)$$
$$b_{ij}=(2k_i^4-k_i^2k_j^2+2k_j^4)\kappa^2-6(k_i^2+k_j^2), (i, j=1, 2, ..., N)
 \eqno(3.21c)$$
$$d_{ij}=(k_i+k_j)^2[(k_i^2+k_ik_j+k_j^2)\kappa^2-3], (i, j=1, 2, ..., N).\eqno(3.21d)$$
The first expression of (3.21a) is a simple consequence of  (2.11d), (2.11e) and (3.2d) while
the second expression follows by substituting (2.11f) and (2.11g) into the first one.
Thus, $L_{m,n}$ becomes 
$$L_{m,n}=c_{m,n}\sum_{\sigma=\pm 1}\prod_{i,j=1\atop(i<j)}^n\left[{1\over 2}
(b_{ij}+\sigma_i\sigma_j\beta_i\beta_j)\right] \eqno(3.22a)$$
where $c_{m,n}$ is a factor independent of  $\sigma_i (i=1, 2, ..., n)$ given by
$$c_{m,n}={{\rm  exp}\left[\sum_{i=n+1}^m
\sum_{j=1\atop(j\not=i)}^m\gamma_{ij}\right]\over 
\prod_{i=1}^na_i\prod_{i,j=1\atop(i<j)}^nd_{ij}}. \eqno(3.22b)$$
Similarly, invoking the relations
$${\rm e}^{-\sigma_i\phi_i}={1\over a_i}{1-{\kappa\over 2}\sigma_ik_i
\over 1+\kappa\sigma_ik_i}, (i=1, 2, ..., N)  \eqno(3.23a)$$
$${\rm exp}\Bigg[{1\over 2} (1+\sigma_i\sigma_j)\gamma_{ij}\Bigg]
={(\sigma_ik_i-\sigma_jk_j)^2
[(k_i^2-\sigma_i\sigma_jk_ik_j+k_j^2)\kappa^2-3] 
\over (k_i+k_j)^2[(k_i^2+k_ik_j+k_j^2)\kappa^2-3]}, $$
$$(i, j=1, 2, ..., N; i\not=j)  \eqno(3.23b)$$
which follow from (2.11c), (3.2c) and (3.2d), we  can modify $R_{m,n}$ in the form
$$R_{m,n}=c_{m,n}\sum_{\sigma=\pm 1}(1+\kappa\sum_{i=1}^n\sigma_ik_i)
\prod_{i=1}^n{1-{\kappa\over 2}\sigma_ik_i \over 1+\kappa\sigma_ik_i}$$
$$\times \prod_{i,j=1\atop(i<j)}^n(\sigma_ik_i-\sigma_jk_j)^2
[(k_i^2-\sigma_i\sigma_jk_ik_j+k_j^2)\kappa^2-3]. \eqno(3.24)$$
Define $P_n$ by the relation 
$$R_{m,n}-L_{m,n}=c_{m,n}P_n(k_1,k_2,...,k_n)  \eqno(3.25a)$$
where
$$P_n(k_1,k_2,...,k_n)=\sum_{\sigma=\pm 1}\left(1+\kappa\sum_{i=1}^n\sigma_ik_i\right)
\prod_{i=1}^n{1-{\kappa\over 2}\sigma_ik_i \over 1+\kappa\sigma_ik_i}$$
$$\times \prod_{i,j=1\atop(i<j)}^n(\sigma_ik_i-\sigma_jk_j)^2
[(k_i^2-\sigma_i\sigma_jk_ik_j+k_j^2)\kappa^2-3]$$
$$-\sum_{\sigma=\pm 1}\prod_{i,j=1\atop(i<j)}^n\left[{1\over 2}
(b_{ij}+\sigma_i\sigma_j\beta_i\beta_j)\right]. \eqno(3.25b)$$
To proceed, we make an important remark.  Let $P_{1,n}$ be the first term
on the right-hand side of (3.25b) multiplied by a factor $\prod_{i=1}^n(1-\kappa^2k_i^2)$.
By virtue of the summation with respect to
$\sigma_i (i=1, 2, ..., n)$, we can see that 
$P_{1,n}$ is symmetric and even function of $k_i (i=1, 2, ..., n)$ and
$P_{1,n}|_{k_1=\pm 1/\kappa}=0$.  Hence,  $P_{1,n}$ has a factor 
$\prod_{i=1}^n(k_i^2-1/\kappa^2)$
in view of symmetry.  This implies that the first term is not a rational function but a polynomial.
Furthermore,  since the second term on the right-hand side of (3.25b) is an even function of
$\beta_i (i=1, 2, ..., n)$ due to the summation with respect to $\sigma_i$, 
only the even power of $\beta_i$ contributes  to the
summation.  This fact indicates that the
second term is indeed a polynomial of  $k_i (i=1, 2, ..., n)$. 
As a consequence, $P_n$ becomes a polynomial of $k_i (i=1, 2, ..., n)$. 
\par
 With the above remark in mind, we now prove 
the identity $P_n=0 (n=1, 2, ..., N)$ by mathematical induction.
A direct calculation shows that $P_1=P_2=0$. We assume that $P_{n-2}=P_{n-1}=0$.
Then
$$P_n|_{k_1=0}=2\prod_{i=2}^nk_i^2(\kappa^2k_i^2-3)P_{n-1}(k_2,k_3, ..., k_n) 
\eqno(3.26a)$$
$$P_n|_{k_1=\pm 2/\kappa}
=2\prod_{i=2}^n[(\kappa^2k_i^2-1)(\kappa^2k_i^2-4)/\kappa^2]
P_{n-1}(k_2,k_3, ..., k_n)  \eqno(3.26b)$$
$$P_n|_{k_1=k_2}=6k_1^2(\kappa^2k_1^2-4)\prod_{i=3}^n(k_i^2-k_1^2)^2
\{(k_i^4+k_1^2k_i^2+k_1^4)\kappa^4-6(k_i^2+k_1^2)\kappa^2+9\}$$
$$\times P_{n-2}(k_3,k_3, ..., k_n).  \eqno(3.26c)$$
On account of the properties (3.26) as well as symmetry 
and evenness, we see that   $P_n$ can be
factored by a polynomial
$$\prod_{i=1}^nk_i^2\left(k_i^2-{4\over \kappa^2}\right)\prod_{i,j=1\atop(i<j)}^n
(k_i^2-k_j^2)^2$$
 of $k_i (i=1, 2, ..., n)$ of degree $2n^2+2n$.
 On the other hand,  it is obvious from (3.25b) that 
$P_n$ is a polynomial of $k_i (i=1, 2, ..., n)$ of
degree $2n(n-1)+1$ at most, which is impossible except for $P_n\equiv 0$.
This completes the proof. \par
\leftline{\it 3.2.2.  Proof of (3.14)}\par
\noindent  The proof of  the identity (3.14) can be done in the same way. By comparing the factor 
${\rm exp}\bigl[\sum_{i=1}^n\xi_i$
\noindent $+\sum_{i=n+1}^m2\xi_i\Bigr]  (0\leq n<m\leq N)$ 
on both sides of (3.14), we obtain
$$\sum_{\sigma=\pm 1}{\rm exp}\Bigg[\sum_{i=1}^n\sigma_i\ {\rm ln}\left(-{q_i\over p_i}\right)-\sum_{i=1}^n{\rm ln}\  a_i
+{1\over 4} \sum_{i,j=1\atop(i\not=j)}^n(1+\sigma_i\sigma_j)\gamma_{ij}
-{1\over 2} \sum_{i,j=1\atop(i\not=j)}^n\sigma_i\sigma_jA_{2i-1,2j}$$
 $$+\sum_{i=n+1}^m\sum_{j=1\atop(j\not=i)}^m\gamma_{ij} \Bigg]
= \sum_{\sigma=\pm 1}{\rm exp}\Bigg[-\sum_{i=1}^n\sigma_i\phi_i
+{1\over 4} \sum_{i,j=1\atop(i\not=j)}^n(1+\sigma_i\sigma_j)\gamma_{ij}
+\sum_{i=n+1}^m\sum_{j=1\atop(j\not=i)}^m\gamma_{ij}\Bigg].
 \eqno(3.27)$$
The calculation leading to (3.25) is applied as well to modify (3.27) further. With
the aid of the relation
$$\left(-{q_i\over p_i}\right)^{\sigma_i}={1+{1\over2}\kappa^2k_i^2-{\kappa\over 2}\sigma_i\beta_i\over
1-\kappa^2k_i^2}, (i=1, 2, ..., N)  \eqno(3.28)$$
which is derived simply from (2.11f) and (2.11g),
we can recast  (3.27) into the form
$$Q_n(k_1,k_2, ..., k_n)\equiv 
 \sum_{\sigma=\pm 1}\prod_{i=1}^n\left(1+{1\over 2}\kappa^2k_i^2-{3\over 2}\kappa\sigma_ik_i\right)$$
 $$\times\prod_{i,j=1\atop(i<j)}^n(\sigma_ik_i-\sigma_jk_j)^2
[(k_i^2-\sigma_i\sigma_jk_ik_j+k_j^2)\kappa^2-3]$$
$$-\sum_{\sigma=\pm 1}\prod_{i=1}^n\left(1+{1\over 2}\kappa^2k_i^2-{1\over 2}\kappa\sigma_i\beta_i\right)
\prod_{i,j=1\atop(i<j)}^n\left[{1\over 2}
(b_{ij}+\sigma_i\sigma_j\beta_i\beta_j)\right]=0. \eqno(3.29)$$
Note that the second term on middle line in (3.29) is an even function 
of  $\beta_i (i=1, 2, ..., n)$ so that
$Q_n$ becomes a polynomial of $k_i (i=1, 2, ..., n)$. 
We now prove  the identity (3.29) by mathematical induction. 
The identity holds for $n=1, 2$, as checked easily. We assume that $Q_{n-2}=Q_{n-1}=0$. 
Then
$$Q_n|_{k_1=0}=2\prod_{i=2}^nk_i^2(\kappa^2k_i^2-3)Q_{n-1}(k_2,k_3, ..., k_n) 
\eqno(3.30a)$$
$$Q_n|_{k_1=\pm 2/\kappa}
=6\prod_{i=2}^n[(\kappa^2k_i^2-1)(\kappa^2k_i^2-4)/\kappa^2]
Q_{n-1}(k_2,k_3, ..., k_n)  \eqno(3.30b)$$
$$Q_n|_{k_1=k_2}=6k_1^2(\kappa^2k_1^2-4)(\kappa^2k_1^2-1)^2
\prod_{i=3}^n(k_i^2-k_1^2)^2
\{(k_i^4+k_1^2k_i^2+k_1^4)\kappa^4-6(k_i^2+k_1^2)\kappa^2+9 \}$$
$$\times Q_{n-2}(k_3,k_3, ..., k_n).  \eqno(3.30c)$$
The symmetry and evenness of $Q_n$ with respect to $k_i(i=1, 2, ..., n)$ as well as the properties (3.30) imply that $Q_n$ has  a factor 
$$\prod_{i=1}^nk_i^2\left(k_i^2-{4\over \kappa^2}\right)\prod_{i,j=1\atop(i<j)}^n
(k_i^2-k_j^2)^2$$
whose degree is $2n^2+2n$. But,  as seen from (3.29), the degree of $Q_n$ is $2n^2$ at most
and hence $Q_n$ must vanish identically, completing the proof.  \par
\leftline{\it 3.3. Proof of the N-soliton solution} \par
\noindent Once the identities (3.13) and (3.14) have been 
established, the proof of the $N$-soliton
solution can be done straightforwardly. Indeed, if we substitute (3.12) with (3.13)
and (3.14) into (2.14) and integrate it with respect to $y$, we obtain (3.1b).
The expression (3.1a) follows immediately from  (3.1b) by
a simple differentiation with respect to $t$ while using (2.5b) and the
constancy of $d$. \par
\leftline{\it 3.4. Remark} \par
\noindent The structure of the $N$-soliton solution thus obtained is worth elucidating. 
The present analysis shows that  the  
$\tau$-functions $g_1$ and $g_2$ defined by (3.2) characterize completely the
$N$-soliton solution of the DP equation. 
Another  $\tau$-function $f$ introduced in (2.11) is then 
expressed in terms of $g_1$ and $g_2$ as  indicated by the key identity (3.13).
It is interesting to recall that  $g_1$ and $g_2$ have the same structure 
as the  $\tau$-function for the
$N$-soliton solution of a model equation for shallow water waves introduced by Hirota
and Satsuma [8]  which is a member of the BKP hierarchy in view of the
classification theory of soliton equations [12, 13]. 
The identity (3.13) shows that the CKP  $\tau$-function $f$ can be represented by the BKP
 $\tau$-functions $g_1$ and $g_2$.  \par
Although we have addressed equation (1.1) with a nonzero $\kappa$, the case
$\kappa=0$ is an interesting problem to call a special attention. As already mentioned in
introduction, the search  dealing with this issue has already been done by several authors [2-6].
In particular, an important problem associated with nonanalytic 
 peakon solutions [3, 5] is 
how to reduce them
from analytic soliton solutions. 
By taking a limit $\kappa\rightarrow  0$, 
a single peakon solution has been reproduced from the one-soliton solution. 
However, the general $N$-peakon case is still left as an open problem  even though we have
a numerical evidence  showing  that  the two-soliton solution with very small $\kappa$ 
approximates quite well the two-peakon solution. 
See  section 4.3 of I. \par
Another limit $\kappa\rightarrow \infty$ will also deserve remarking. It has been
shown that  if one
replaces the velocity of the $i$th soliton by $k_i^4$ in (2.11b) and  takes the 
limit $\kappa\rightarrow \infty$, then
the  $\tau$-function reduced from $f$ given by (2.11a) provides the $N$-soliton solution
of the following fifth-order KdV equation introduced by Kaup [14] 
which belongs to a member of the CKP hierarchy [13, 15]
$$v_t+\left(v_{yyyy}+20vv_{yy}+{80\over 3}v^3+15v_y^2\right)_y=0.  \eqno(3.31)$$
To clarify the structure of the limiting form of the $\tau$-function $f$, it is suitable to
use (3.2) and (3.13). We find in the limit of 
$\kappa\rightarrow \infty$ that
$g_1$ and $g_2$ are expanded in inverse power of $\kappa$ as
$$g_1 \sim h-{1\over \kappa}h_{t^\prime}+O\left({1\over\kappa^2}\right)  \eqno(3.32a)$$
$$g_2 \sim h+{1\over \kappa}h_{t^\prime}+O\left({1\over\kappa^2}\right) 
 \eqno(3.32b)$$
where 
$$h=\sum_{\mu=0,1}{\rm exp}\left[\sum_{i=1}^N\mu_i\hat\xi_i
+\sum_{i,j=1\atop (i<j)}^N\mu_i\mu_j\hat\gamma_{ij}\right]
\eqno(3.33a)$$
$$\hat\xi_i=k_i\left(y-k_i^4t+{3\over k_i^2}t^\prime-y_{i0}\right), 
(i=1, 2, ..., N)  \eqno(3.33b)$$
$${\rm e}^{\hat\gamma_{ij}}={(k_i-k_j)^2(k_i^2-k_ik_j+k_j^2)\over
 (k_i+k_j)^2(k_i^2+k_ik_j+k_j^2)}, (i, j=1, 2, ..., N; i\not=j). \eqno(3.33c)$$
Here, $t^\prime$ is an auxiliary time variable 
which may be either set to zero or absorbed into 
the phase constant $y_{i0}$ in the final stage 
of the calculation.  Substitution of (3.32) into (3.13) yields
$$f \sim \hat f\equiv h^2-2(hh_{t^\prime y}-h_{t^\prime}h_y)  \eqno(3.34)$$
and $v$ with $f=\hat f$  in (2.10) gives the $N$-soliton solution of  equation (3.31). 
Hence, a single  $\tau$-function $h$ characterizes completely the $N$-soliton structure.
See also  [16-18] as for the construction of  soliton solutions to equation (3.31) by
means of the bilinear approach. \par
\bigskip
\leftline{\bf 4. Asymptotic behavoir of the $N$-soliton solution}\par        
\noindent  The asymptotic form of the $N$-soliton solution has been investigated in I.
Here, we provide an alternative but more straightforward way to derive it 
 on the basis of the asymptotic form of the $\tau$-functions $g_1$ and $g_2$. 
To proceed, we order the magnitude of the soliton velocity in the $(x, t)$ coordinate system as
$$c_1>c_2> ...>c_N  \eqno(4.1a)$$
with
$$c_i={3\kappa^3\over 1-\kappa^2k_i^2}, (i=1, 2, ..., N)  \eqno(4.1b)$$
and then transform to a moving reference frame  with a constant
velocity $c_i$.
We first take the limit $t\rightarrow -\infty$ with the phase variable 
$\xi_i$ of the $i$th soliton being fixed. Since other phase variables
tend to $\pm\infty$ as 
$$\xi_1, \xi_2, ..., \xi_{i-1}\rightarrow +\infty,
 \ \xi_{i+1}, \xi_{i+2}, ..., \xi_{N}\rightarrow -\infty  \eqno(4.2)$$
the  $\tau$-functions $g_1$ and $g_2$ given respectively by  (3.2a) 
and (3.2b) are found to have the leading-order
asymptotics
$$g_1 \sim {\rm exp}\left[{\sum_{j=1}^{i-1}(\xi_j-\phi_j)}\right]
\left(1+{\rm e}^{\xi_i-\phi_i  +\gamma_i^{(+)}}\right)  \eqno(4.3a)$$
$$g_2 \sim {\rm exp}\left[{\sum_{j=1}^{i-1}(\xi_j+\phi_j)}\right]
\left(1+{\rm e}^{\xi_i+\phi_i  +\gamma_i^{(+)}}\right)  \eqno(4.3b)$$
where
$$\gamma_i^{(+)}=\sum_{j=1}^{i-1}\gamma_{ij}, (i=1, 2, ..., N).
\eqno(4.3c)$$
The asymptotic forms of $u$ and $x$ follow simply by substituting (4.3)
into (3.1).  They  can be written as
$$u \sim u_i(\xi_i+\gamma_i^{(+)})  \eqno(4.4a)$$
$$x-c_it-x_{i0} \sim {\xi_i\over \kappa k_i}
+{\ln}\left({1+{\rm e}^{\xi_i-\phi_i  +\gamma_i^{(+)}}
          \over 1+{\rm e}^{\xi_i+\phi_i  +\gamma_i^{(+)}}}\right)
-2\sum_{j=1}^{i-1}\phi_j +d  \eqno(4.4b)$$
where  $u_i(\xi_i)$ is the
one-soliton solution
$$u_i(\xi_i)={\kappa k_i c_i\sinh\ \phi_i
\over \cosh\ \xi_i+\cosh\ \phi_i} \eqno(4.4c)$$
and $x_{i0}=y_{i0}/\kappa$.
We can confirm that the above expression of $u_i$ 
coincides with the one-soliton solution given by  (2.15a).
Actually, by virtue of  the relations
$$\cosh\ \phi_i = 2a_i-{1\over a_i}  \eqno(4.5a)$$
$$\sinh\ \phi_i = {1\over a_i}\sqrt{(4a_i^2-1)(a_i^2-1)}  \eqno (4.5b)$$
$$\kappa k_i = \sqrt{a_i^2-1\over a_i^2-{1\over 4}}  \eqno(4.5c)$$
which come from (2.11c) and (3.2c), we can rewrite (4.4c) as
$$u_i(\xi_i)={{8\kappa^3\over a_i}(a_i^2-1)\left(a_i^2-{1\over 4}\right)
\over \cosh\ \xi_i +2a_i-{1\over a_i}}  \eqno(4.6)$$
which is just (2.15a). 
\par
In the limit of $t \rightarrow +\infty$, on the other hand, the
expressions corresponding to (4.4a) and (4.4b) turn out to be as
$$u \sim u_i(\xi_i+\gamma_i^{(-)})  \eqno(4.7a)$$
$$x-c_it-x_{i0} \sim {\xi_i\over \kappa k_i}
+{\ln}\left({1+{\rm e}^{\xi_i-\phi_i  +\gamma_i^{(-)}}
          \over 1+{\rm e}^{\xi_i+\phi_i  +\gamma_i^{(-)}}}\right)
-2\sum_{j=1}^{i-1}\phi_j +d  \eqno(4.7b)$$
 where
$$\gamma_i^{(-)}=\sum_{j=i+1}^{N}\gamma_{ij}, (i=1, 2, ..., N).
\eqno(4.7c)$$
 If one observes the behavior of the $N$-soliton solution described
above in the coordinate system at rest, the asymptotic form of the
solution as $t\rightarrow \pm\infty$ is represented by a superposition of $N$  single
solitons
$$u \sim \sum_{i=1}^Nu_i(\xi_i+\gamma_i^{(+)}), (t \rightarrow -\infty)
\eqno(4.8a)$$
$$u \sim \sum_{i=1}^Nu_i(\xi_i+\gamma_i^{(-)}), (t \rightarrow +\infty).
\eqno(4.8b)$$
Note that if one shifts the phase $\xi_i$ as $\xi_i\rightarrow \xi_i-\sum_{j=1\atop(j\not=i)}^N
\gamma_{ij}$, then the argument of $u_i$ in (4.8a) and (4.8b) becomes 
 $ \xi_i-\gamma_i^{(-)}$ and $\xi_i-\gamma_i^{(+)}$, respectively. The 
resulting asymptotic expressions of $u$ coincide with those derived in I 
by a different method (see  formulas (4.45) and (4.50) of I). 
It should be emphasized, however, that these asymptotic formulas are valid
in the $(y, t)$ coordinate system. The new feature appears when we 
transform back to the original $(x, t)$ coordinate system, which we shall
now demonstrate.
Indeed, as $t\rightarrow -\infty$, the center position of the $i$th
soliton in the $(y, t)$ coordinate is found from (4.8a) 
as $\xi_i=-\gamma_i^{(+)}$. It then follows from (4.4b) that
the trajectory of the corresponding center position $x_c$ in 
the $(x, t)$ coordinate is described by  the relation
$$ x_c-c_it-x_{i0} = -{\gamma_i^{(+)}\over \kappa k_i}
+{\rm ln}\left({1+{\rm e}^{-\phi_i}\over 1+{\rm e}^{\phi_i}}\right)
-2\sum_{j=1}^{i-1}\phi_j +d.
\eqno(4.9)$$
As $t\rightarrow +\infty$, the expression corresponding to (4.9)
 takes the form
 $$ x_c-c_it-x_{i0} = -{\gamma_i^{(-)}\over \kappa k_i}
+{\rm ln}\left({1+{\rm e}^{-\phi_i}\over 1+{\rm e}^{\phi_i}}\right)
-2\sum_{j=i+1}^{N}\phi_j +d.
\eqno(4.10)$$
  Let $\Delta_i$ be the phase shift of the $i$th soliton which is
defined by the shift of the trajectory of the $i$th soliton as $t\rightarrow +\infty$ 
relative to its trajectory as  $t\rightarrow  -\infty$.  This quantity is evaluated  simply
 from (4.9) and (4.10), giving rise to the result
$$\Delta_i={1\over \kappa k_i}\left(\gamma_i^{(+)}-\gamma_i^{(-)}\right)
+2\sum_{j=1}^{i-1}\phi_j-2\sum_{j=i+1}^{N}\phi_j,
(i=1, 2, ..., N). \eqno(4.11)$$
If we use the formula
$${\rm e}^{-\phi_i}={\alpha_i-1\over \alpha_i+1},  
(i=1, 2, ..., N)  \eqno(4.12a)$$
with
$$\alpha_i=\sqrt{(2a_i-1)(a_i+1)\over (2a_i+1)(a_i-1)},
(i=1, 2, ..., N)  \eqno(4.12b)$$
which is derived from (2.11c) and (3.2c), formula (4.11)
reproduces  the corresponding expression  presented  in  I (see (4.52) of I).
While the first term on the right-hand side of (4.11) is in accordance with
the formula for the phase shift arising in the context of the
shallow water wave equation [8],  the second and third terms appear
as  a consequence  of the coordinate transformation (2.3).
In view of the latter terms, the characteristics of the
interaction process of solitons are found to differ substantially from
those derived from the shallow water theory. 
A full explanation about this topic  has already been given
  in I in the case of the two-soliton solution.  \par
\bigskip
\leftline{\bf 5. Discussion} \par
\noindent In conclusion, it will be worthwhile to discuss the structure of the $N$-soliton 
solution of the DP equation in conjunction with that of the following CH
equation [19-21]
$$u_t+2\kappa^2 u_x-u_{xxt}+3uu_x=2 u_xu_{xx}+uu_{xxx}. \eqno(5.1)$$
The multisoliton solutions of the CH equation with  $\kappa\not=0$ have
been constructed by several different methods [9, 22-27].
In particular, we have  demonstrated by using an
elementary theory of determinants that the $N$-soliton solution of the CH equation exhibits
a parametric representation similar to the $N$-soliton solution (3.1) of the
DP equation. Indeed, it can be written as [9]
$$u(y,t)=\left({\rm ln}\ {f_2\over f_1}\right)_t  \eqno(5.2a)$$
$$x(y,t)={y\over\kappa} +{\rm ln}\ {f_2 \over f_1} +d, \eqno(5.2b)$$
where $f_1$ and $f_2$ are  $\tau$-functions whose structure 
is essentially the same as 
that of the $\tau$-functions 
 constructing the $N$-soliton solution of a model equation for 
shallow water waves introduced in the context of the inverse scattering 
transform (IST) method [26]. Thus, the difference between the  $\tau$-functions
$g_1, g_2$ and $f_1, f_2$ turns out to be the most important issue to be explored.
Although $f_1$ and $f_2$  have the determinantal expressions [9],  $g_1$ and $g_2$
are expressed  by a finite sum as indicated in (3.2).  An  important observation is
that $g_1$ and $g_2$ can be written in terms of pfaffians [12, 13, 29]. 
These  $\tau$-functions
satisfy the bilinear equation [8] for a model shallow water wave equation 
of Hirota and Satsuma
which is a member of the BKP hierarchy. On the other hand, it is well known that
$f_1$ and $f_2$ satisfy the bilinear equation [8, 25] for a model shallow water wave equation
of Ablowitz et al which is a member of  the AKP hierarchy.  
These inspections show clearly 
that the DP and CH equations
belong to the different class of soliton equations.  The discussion given here becomes more
transparent if one compares the Lax pairs associated with both equations. 
Indeed, the spectral problem associated with the DP equation is described 
by the third order equation [2]  while the one corresponding to the CH equation is the
second order equation [20, 21]. As far as we know, however,
the former spectral problem has not been 
studied sufficiently as yet.  In this respect, it is an interesting problem to derive the $N$-soliton
solution of the DP equation by means of IST. 
\par
\bigskip
\leftline{\bf  Acknowledgement}\par
\noindent This work was partially supported by the Grant-in Aid for
Scientific Research (C) No. 16540196 from the Ministry of Education, Culture,
Sports, Science and Technology. \par
\bigskip
\leftline{\bf Appendix. The CKP  $\tau$-function} \par
\noindent We introduce the KP $\tau$-function in the standard Hirota form [30]
$$\tau_{2N}=\sum_{\mu=0,1}{\rm exp}
\left[\sum_{i=1}^{2N}\mu_i\tilde\xi_i+\sum_{i,j=1\atop(i<j)}^{2N}
\mu_i\mu_j\tilde A_{ij}\right]  \eqno(A.1)$$
with
$$\tilde\xi_i =(\tilde p_i-\tilde q_i)y
+\left({1\over\tilde p_i}-{1\over\tilde q_i}\right)\kappa^2t
+\tilde\xi_{i0}, (i=1, 2, ..., 2N)  \eqno(A.2)$$
$${\rm e}^{\tilde A_{ij}}={(\tilde p_i-\tilde p_j)(\tilde q_i-\tilde q_j)
    \over (\tilde p_i-\tilde q_j)(\tilde q_i-\tilde p_j)},
(i, j=1, 2, ..., 2N; i\not=j)  \eqno(A.3)$$
where the notation $ \sum_{\mu=0,1}$ implies the summation
over all possible combinations of $\mu_1=0, 1, \mu_2=0, 1, ..., 
\mu_{2N}=0, 1$. 
While  the usual definition of the  $\tau$-function introduces an infinite
number of independent variables, only the two variables $y$ and $t$ are
retained  and others are set to zero  in (A.2).
The CKP reduction is defined by the following
parameterization [15]
$$\tilde p_{2i-1}=q_i, \tilde p_{2i}=p_i, \tilde q_{2i-1}=-p_i, 
\tilde q_{2i}=-q_i, (i=1, 2, ..., N). \eqno(A.4)$$
In order to obtain the $N$-soliton sollution of equation (2.9),  the parameters $p_i$ and $q_i$ 
are specified   by (2.11f) and (2.11g),  respectively.
It then follows from (A.1), (A.4), (2.11f) and (2.11g) that
$$\tilde\xi_{2i-1}=\tilde\xi_{2i}=\xi_i,
(i=1, 2, ..., N). \eqno(A.5)$$
Here, $\xi_i$ is given by (2.11b). Note that the phase
factors $\tilde\xi_{i0}\ (i=1, 2, ..., 2N)$ have been 
rewritten as $\tilde\xi_{2i-1,0}=\tilde\xi_{2i,0}=(p_i+q_i)y_{i0}
\ (i=1, 2, ..., N)$. We denote the CKP  $\tau$-function  constructed above
by $\tau_{2N}|_{CKP}$. We have already shown in I that the function 
$v\equiv (3/4)({\rm ln}\ f)_{yy}$ with $f=\tau_{2N}|_{CKP}$ (see (2.10))
satisfies equation (2.9).  We shift the phase of the $i$th soliton as
$\xi_i\rightarrow \xi_i-{\rm ln}\ a_i$ (or equivalently, 
$y_{i0}\rightarrow y_{i0}+({\rm ln}\ a_i/k_i)$) and rewrite $f$ in the form
$$f=\sum_{\mu=0,1}{\rm exp}\Biggl[
\sum_{i=1}^N(\mu_{2i-1}+\mu_{2i})(\xi_i-{\rm ln}\ a_i)$$
$$+{1\over 2}\sum_{i,j=1\atop(i\not=j)}^N
(\mu_{2i-1}\mu_{2j-1}A_{2i-1,2j-1}+\mu_{2i}\mu_{2j}A_{2i,2j})$$
$$+{1\over 2}\sum_{i,j=1\atop(i\not=j)}^N
(\mu_{2i-1}\mu_{2j}A_{2i-1,2j}+\mu_{2i}\mu_{2j-1}A_{2i,2j-1})$$
$$+{1\over 2}\sum_{i=1}^N
(\mu_{2i-1}\mu_{2i}A_{2i-1,2i}+\mu_{2i}\mu_{2i-1}A_{2i,2i-1})\Biggr] 
\eqno(A.6)$$
where $A_{2i-1,2j-1}(=A_{2i,2j})$ and $A_{2i-1,2j}(=A_{2i,2j-1})$ 
are given respectively by (2.11d) and (2.11e). We replace the
dummy indices $\mu_i (i=1, 2, ..., 2N)$ in accordance with the rule
$\mu_{2i-1}\rightarrow \mu_i, 
\mu_{2i}\rightarrow \nu_i (i=1, 2,..., N)$. If we substitute the
relations $A_{2i-1,2i}=A_{2i,2i-1}=2\ {\rm ln}\ a_i$ which can be derived
from  (2.11e), (2.11f) and (2.11g), then we arrive at 
formula (2.11a).  \par
In the present analysis, we have obtained the  $\tau$-function in the form of finite sum.
We can also express it in terms of a determinant, as already demonstrated in I. 
To see the relationship between two alternative expressions of $f$, we use formula (3.7) with
the phase variables $\xi_i$ shifted appropriately.  In the two-soliton case, for instance, we
shift $\xi_i$ as $\xi_i\rightarrow \xi_i-{\rm ln}\ \delta (i=1, 2)$  in (2.12b) and then
multiply  it by a constant factor $(a_1a_2\delta)^2$.  We can confirm 
that the resultant expression
coincides with the corresponding  $\tau$-function presented in I (see (4.18) of I). \par
\vfill\eject
{\bf References}\par
\item{[1]}  Degasperis A and  Procesi M 1999   {\it Symmetry and Perturbation Theory},
ed. A Degasperis and G Gaeta (Singapore: World Scientific ) pp 22-37
\item{[2]}  Degasperis A, Hone ANW and  Holm DD 2002 
{\it Theor. Math. Phys.} {\bf 133} 1463-74  \par
\item{[3]}  Lundmark H and  Szmigielski J 2003  {\it Inverse Problems} {\bf 19} 
1241-5  \par
\item{[4]} Vakhnenko VO  and  Parkes EJ  2004   
{\it Chaos, Solitons and Fractals} {\bf 20} 1059-73  \par
\item{[5]} Lundmark H  and  Szmigielski J 2005  {\it Internat. Math. Res. Paper} 
no 2  53-116 \par
\item{[6]} Guo B and  Liu Z 2005  {\it Chaos, Solitons and Fractals}
{\bf 23} 1451-63  \par
\item{[7]} Matsuno Y 2005  {\it Inverse Problems} {\bf 21} 1553-70 \par
\item{[8]} Hirota R and  Satsuma J 1976  {\it J. Phys. Soc. Jpn.} {\bf 40} 611-3 \par
\item{[9]} Matsuno Y 2005  {\it J. Phys. Soc. Jpn.} {\bf 74} 1983-7 \par
\item{[10]} Matsuno Y 2000 {\it Phys. Lett. }{\bf A 278} 53-8 \par
\item{[11]} Hirota R and Satsuma J 1976 {\it Prog. Theor. Phys. Suppl.} {\bf 59} 64-100
\par
\item{[12]}  Date E,  Jimbo M,  Kashiwara M and  Miwa T 1982 {\it Physica }{\bf 4D}  343-65  \par
\item{[13]}  Jimbo M and  Miwa T 1983 {\it Publ. RIMS Kyoto Univ.} {\bf 19}  943-1001 \par
\item{[14]}  Kaup DJ 1980  {\it Stud. Appl. Math.} {\bf 62}  189-216 \par
\item{[15]}  Date E,  Jimbo M,  Kashiwara M and  Miwa T 1981 {\it  J. Phys. Soc. Jpn.} 
{\bf  50} 3813-18 \par
\item{[16]}  Hirota R and  Ramani A 1980 {\it RIMS Kokyuroku} {\bf 375} 129-35 \par
\item{[17]}  Parker A 1990 {\it Physica} {\bf D137} 25-33 \par
\item{[18]}  Parker A 1990 {\it  Physica }{\bf D137} 34-48 \par
\item{[19]}  Fuchssteiner B and  Fokas A 1981 {\it Physica} {\bf D4} 47-66 \par
\item{[20]}  Camassa R and  Holm DD 1993 {\it Phys. Rev. Lett.} {\bf 71}  1661-4 \par
\item{[21]} Camassa R, Holm DD and Hyman J 1994 {\it Adv. Appl. Mech.} {\bf 31}
           1-33 \par
\item{[22]}  Schiff J 1998 {\it  Physica }{\bf  D 121} 24-43 \par
\item{[23]}  Johnson RS 2003 {\it Proc. R. Soc. London Ser. A} {\bf  459}  1687-708 \par
\item{[24]}  Li Y and  Zhang JE  2004 {\it Proc. R. Soc. London Ser. A} 
           {\bf  460}  2617-27 \par
\item{[25]}  Parker A 2004 {\it  Proc. R. Soc. London} Ser. A {\bf  460} 2929-57 \par
\item{[26]}  Li Y 2005 {\it  J. Nonl. Math. Phys. }{\bf 12} Supplement 1 466-81
\item{[27]}  Parker A 2005 {\it  Proc. R. Soc. London Ser. A} {\bf 461} 3611-32, 3893-911    \par
\item{[28]}  Ablowitz MJ,  Kaup DJ, Newell AC and  Segur H 1974 
           {\it Stud. Appl. Math.} {\bf 53} 249-315  \par
\item{[29]}  Hitota R 1989 {\it J. Phys. Soc. Jpn.} {\bf  58} 2285-96  \par
\item{[30]}  Date E,   Kashiwara M and  Miwa T 1981 {\it  Proc. Japan Acad. Ser. A }
          {\bf 57} 387-92 \par
\bye